\documentclass[prd,twocolumn,showpacs,preprintnumbers,nofootinbib,amsmath,amssymb,nobalancelastpage]{revtex4}

\usepackage[pdftex]{graphicx}
\usepackage{latexsym,amsmath,amssymb,lmodern,float,url}
\usepackage{natbib}
\usepackage[pdftex,bookmarks,linktocpage,pdfpagelabels,plainpages=false,hyperfigures,linkcolor=blue,citecolor=blue,urlcolor=blue]{hyperref} 
\usepackage{xspace}
\usepackage{courier}
\hypersetup{colorlinks=true}
\usepackage{bm}
\usepackage{dcolumn}
\usepackage{amsmath}
\usepackage{amssymb}
\usepackage{color}
\usepackage{xcolor}
\usepackage{soul}
\usepackage{url}
\usepackage{float}

\usepackage{aas_macros}
\usepackage[normalem]{ulem}

\def\Eq#1{Eq.~(\ref{#1})}
\def\Fig#1{Fig.~(\ref{#1})}

\def\H0{$H_0$}
\def\si8{$\sigma_8$}

\def\Msun{M_\odot}

\graphicspath{{./figs/}}

\begin{document}

\title{Could the 2.6 $M_\odot$ object in GW190814 be a primordial black hole?}

\author{ Kyriakos Vattis}
\email{kyriakos$\_$vattis@brown.edu}
\affiliation{ Department of Physics, Brown University, Providence, RI 02912-1843, USA}
\affiliation{ Brown Theoretical Physics Center, Brown University, Providence, RI 02912-1843, USA}

\author{Isabelle S. Goldstein} 
\email{isabelle$\_$goldstein@brown.edu}
\affiliation{ Department of Physics, Brown University, Providence, RI 02912-1843, USA}
\affiliation{ Brown Theoretical Physics Center, Brown University, Providence, RI 02912-1843, USA}

\author{Savvas M. Koushiappas}
\email{koushiappas@brown.edu}
\affiliation{ Department of Physics, Brown University, Providence, RI 02912-1843, USA}
\affiliation{ Brown Theoretical Physics Center, Brown University, Providence, RI 02912-1843, USA}

\date{\today}

\begin{abstract}
On June 20, 2020, the LIGO-Virgo collaboration announced the discovery of GW190814, a gravitational wave event originating from a binary system merger between a black hole of mass $M_1 = 23.2^{+1.1} _ {-1.0}M_\odot$ and an unidentified object with a mass of $M_2 = 2.59^{+0.08} _ {-0.09}M_\odot$. This second object would be either the heaviest neutron star or lightest black hole observed to date.  Here we investigate the possibility of the $\sim 2.6M_\odot$ object being a primordial black hole (PBH). We find that a primordial black hole explanation to GW190814 is unlikely as it is limited by the formation rate of the primary stellar progenitor and the time available for a pair of primordial- and stellar-origin black hole binaries to form and merge within a hubble time. 
\end{abstract}

\maketitle

\section{Introduction}

In  recent years high precision cosmological and astrophysical observations have established $\Lambda$CDM as the Standard Cosmological Model \cite{Planck2018VI}. However one of its main components, dark matter, has only been observed through gravitational effects and thus its exact nature remains illusive. Direct and indirect detection experimental searches \cite{FroborgDuffy2020,Gaskins2016, 2017PhRvL.118b1303A, 2017arXiv170808869S, 2012PhLB..709...14A, 2015JCAP...09..008F, 2015PhRvD..91h3535G, 2018ApJ...853..154A, 2017PhRvD..95h2001A, 2020Galax...8...25R,2016JCAP...02..039M,2017arXiv170508103I,2015arXiv150304858T} as well as the Large Hadron Collider \cite{Kahlhoefer2017, 2017Sirunyan, 2019Aaboud} have been unsuccessfully searching for a Weakly Interacting Massive Particle (WIMP) as a dark matter candidate. The most recent observed anomaly detected in the XENON1T experiment does not match the required characteristics \cite{kannike2020dark,2020arXiv200612431V} (however for a possible explanation see \cite{2020arXiv200612488B}), while the parameter space of other popular particle candidates such as axion dark matter is shrinking \cite{2015AxionAnnualReview,PhysRevLett.120.151301}. Other astrophysical candidates alternative to the particle hypothesis such as dark matter in the form of Massive Compact Halo Objects (MACHOs) have been considered, however they are also heavily constrained from microlensing experiments \cite{Alcock_2000, Tisserand2007}.

In light of gravitational wave detections by the LIGO Collaboration \cite{LIGO2019} originating from binary black hole mergers with masses of about $30 M_\odot$ another dark matter candidate possibility resurfaced: primordial black holes (PBHs) formed in the early universe prior to Big Bang Nucleosynthesis \cite{Hawking1971, CarrHawking1974,Carr1975}. The possibility that LIGO has already detected dark matter in the form of $30\Msun$ PBH mergers is investigated in \cite{Bird2016}, and a wealth of other work has been done on the merger rate of PBHs \cite{Gow:2019pok, Young:2019gfc, Vaskonen:2019jpv, Ballesteros:2018swv, Chen:2018czv}. 
These black holes can span numerous orders of magnitude in mass  but, similarly to MACHOs, their parameter space has been heavily constrained \cite{carr2020constraints} though a combination of microlensing at lower masses \cite{Niikura2019, Allsman2001, Wyrzykowski2011, Tisserand2007}, Cosmic Microwave Background (CMB) experiments \cite{poulter2019cmb, 2017PhRvD..95d3534A}, and dynamical effects in Milky Way dwarf galaxies \cite{2017PhRvL.119d1102K, 2016ApJ...824L..31B, 2020MNRAS.492.5247S}.

Despite that, there are still three windows for PBH masses in which PBHs can make up $[1-10 ]\%$ of the dark matter energy budget \cite{carr2020primordial} and thus they remain an interesting option for further investigations, $[10^{16}- 10^{17}]$g, $[10^{20}- 10^{24}]$g, and $[1-10^2] M_\odot$. 
\cite{carr2020constraints} 
However it's important to note that these constraints typically assume a monochromatic PBH mass function as it is possible that a continuum distribution of masses can be established at formation  \cite{PhysRevD.47.4244,PhysRevD.58.107502,PhysRevD.94.063530, PhysRevD.96.023514, PhysRevD.95.083508}. 

Recently the LIGO-Virgo collaboration announced the discovery of GW190814 \cite{Abbott_2020}, a gravitational wave event originating from a binary system merger with a very small ratio of masses $M_2/M_1 = 0.112 ^{+0.008} _ {-0.009}$. The primary component was identified as a black hole of mass $M_1 = 23.2^{+1.1} _ {-1.0}M_\odot$ while the secondary object is unidentified with a mass of $M_2 = 2.59^{+0.08} _ {-0.09}M_\odot$. This event is surprising for two reasons; this is the most asymmetrical binary mass ratio to date and the secondary object's mass lies within the so called ``low mass gap".  This gap between $\sim 2 - 5 M_\odot$ owes to a complete lack of observations, in gravitational and electromagnetic waves, of black holes with mass less than $5 M_\odot$ or neutron stars with mass above $\sim 2 M_\odot$ \cite{Ozel2010,Farr2011,Ozel2012} which are backed by the fact that current accepted stellar evolutionary models are not able to predict compact objects in that mass range, depending on the progenitor explosion mechanism \cite{2012ApJ...749...91F}. Thus if we interpret this event as a new category of binary system mergers the derived merger rate is between $1-23$ Gpc$^{-3}$yr$^{-1}$.

Any attempts to explain this relatively high rate struggle to do so within standard astrophysical and cosmological models.  As discussed in \cite{Abbott_2020} and in more detail in \cite{2020arXiv200614573Z}, this observation challenges most results obtained from population synthesis simulations for isolated binaries. The observed rate cannot be explained by dynamical arguments \cite{2020ApJ...895L..15F} nor a low-mass merger remnant that acquires a BH companion via dynamical interactions in dense environments due to the lack of mass segregation of neutron stars \cite{2020ApJ...888L..10Y}. Other proposals include that this event was subject to gravitational lensing as discussed in \cite{broadhurst2020interpreting}, accretion of  supernova ejecta mass from a neutron star formation that remained bound in a binary system \cite{2020arXiv200700847S}, mergers in wide hierarchical quadruple systems \cite{2020ApJ...888L...3S}. This uncertainty in the predicted rates opens up the possibility that GW190814 is the result of a previously unknown population of mergers. There have also been several analysis considering the possibility that the $2.6 \Msun$ object is a neutron star, with resulting constraints on the neutron star equation of state or exotic degrees of freedom
\cite{godzieba2020maximum, 2020arXiv200706526L, 2020arXiv200705526T, 2020arXiv200703799F, 2020arXiv200616296T, 2020arXiv200708493D}. 

Here, we investigate the possibility that the $2.6\Msun$ object in GW190814 is a primordial black hole. Such an explanation to GW190814 requires knowledge of the merger rate of stellar mass black holes with primordial black holes, while the dynamics of the formation and merger of such binaries is regulated by the formation rate of heavy stellar mass black holes. 

\begin{figure}[ht]
\centering
\includegraphics[width=1.0\columnwidth]{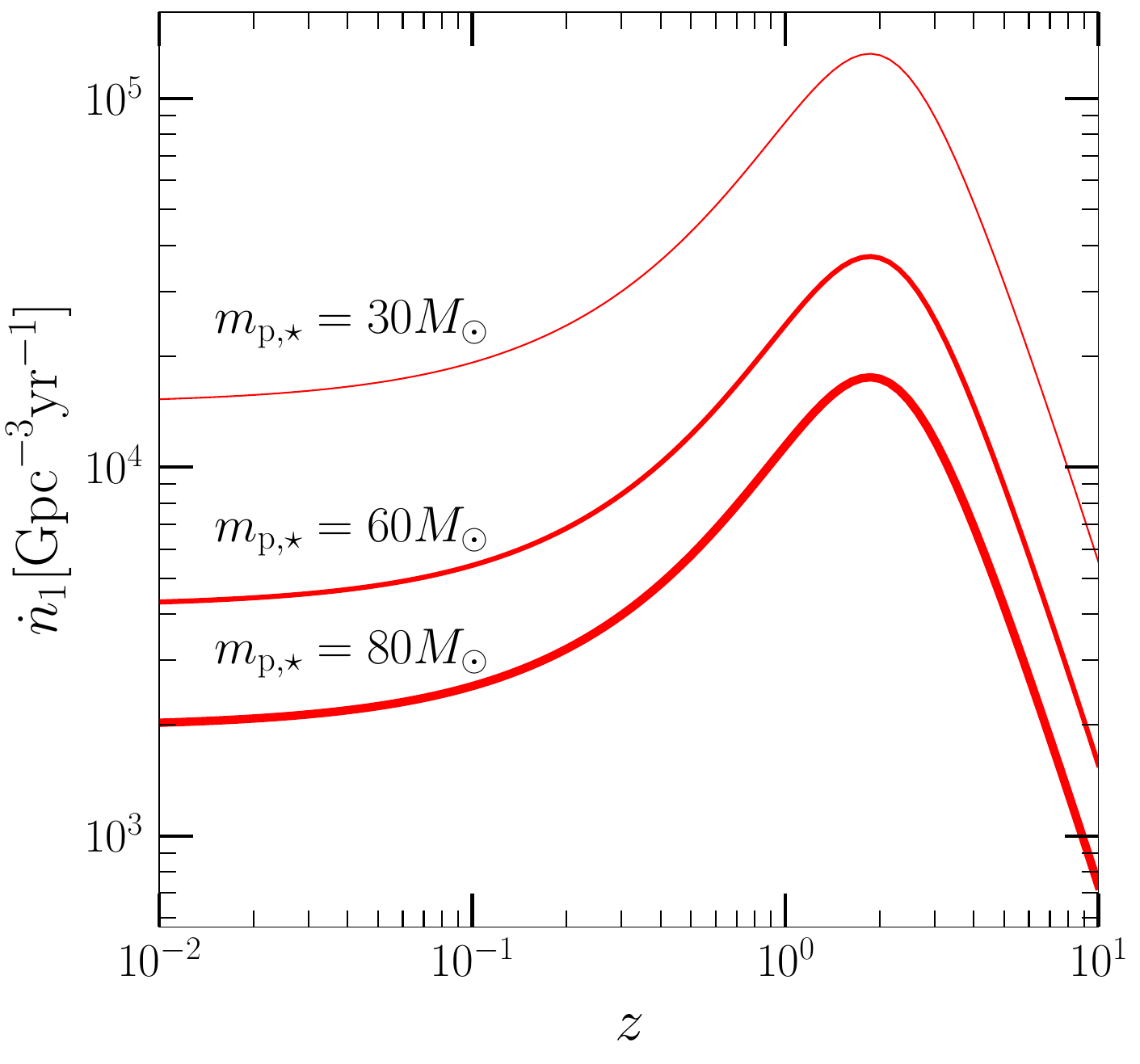} 
\caption{The formation rate of $23\Msun$ black holes, under the maximal assumption that every star with mass greater than $m_{p,*}$ will produce a $23\Msun$ black hole. The thin, medium and thick curves correspond to $m_{p,*} = [30\Msun, 60 \Msun, 80\Msun ]$ respectively.  }
\label{fig:fig1}
\end{figure}

\section{primordial black hole -- stellar black hole merger rate}

We would like to characterize the probability of a $2.6\Msun$ primordial black hole merger with a $23 \Msun$ stellar remnant. Our plan is to first describe the rate density for such an event and by comparing it with the observed rate density obtained from GW190814 we will be able to assess its probability. In what follows, we will use `2' as a subscript that denotes primordial black holes (e.g., of mass $2.6\Msun$), and `1'  a subscript that denotes a compact stellar remnant (e.g., a black hole of mass  $23 \Msun$). 

The merger rate between a primordial black hole and a stellar remnant can then be written as 
\begin{equation} 
\mathcal{R}= \int {\cal{P}} \, \dot{n}_1(z) \, n_2(z) \,   \frac{dV}{dz} \frac{dz}{1+z}. 
\label{eq:merger_rate}
\end{equation}
The rate as given in \Eq{eq:merger_rate} has units of merger events per year per volume, $n_2$ corresponds to the number density of primordial black holes, $\dot{n}_1$ is the number density of stellar remnant formation at $z$ per unit time, and $dV/dz$ is the cosmological volume element. The factor of $1 +z $ in the denominator ensures proper conversion between comoving and physical time intervals. 

The quantity $ {\cal{P}}$ describes the probability of such a merger to occur. In some ways one may think of $ {\cal{P}}$ as a dimensionless ``cross section" for such an interaction. It encapsulates all the assumptions and uncertainties that stem from our lack of knowledge of the precise physics that drives such a processes. For example a primordial black hole and a stellar origin black hole will become bound if during weak gravitational scattering the energy loss brings the total energy of the system below the initial kinetic energy of the pair. This process depends on the number density and velocity distribution of black holes. Once the pair is bound, it will take some time, $\tau$, for gravitational wave emission to dissipate the orbital energy of the system and lead to the merger of the two black holes. If during that time, a third black hole interacts with the system then the binary will harden faster with the ejection of the lightest black hole. The quantity $ {\cal{P}}$ in \Eq{eq:merger_rate} qualitatively captures the net probability of the merger, and therefore can be used to assess the potential of a merger between a $2.6 \Msun$ primordial black hole and a $23 \Msun$ black hole of stellar origin.

We will now describe each term that enters in the rate calculation. The number density of primordial black holes of mass $M_2$ at redshift $z$ can be expressed as 
\begin{eqnarray} 
n_2(z) &=& f_2 \frac{\rho_{\mathrm{DM}}}{ M_2} \\
	&=& f_2 \Omega_M(1 + z)^3 \frac{\rho_{\mathrm{crit}}}{M_2}, 
	\label{eq:n1}
\end{eqnarray} 
With $\Omega_M$ taken to be $\Omega_M = 0.3$. The quantity $f_2$ is the fraction of the dark matter density in the form of primordial black holes. This fraction is heavily constrained by a swarm of observational arguments \cite{Niikura2019, carr2020constraints, poulter2019cmb, carr2020primordial}, but in general, around $M_2 \approx 2.6 \Msun$, $f_2$ is less than 1\%. 

It is important to note that by writing \Eq{eq:n1} in this fashion we make the implicit assumption  of a monochromatic distribution of primordial black hole masses (of $M_2 \approx 2.6 \Msun$). If instead we assume a spectrum of masses then the abundance at any given mass will be lower as compared to the monochromatic case. In addition, the number density of primordial black holes in \Eq{eq:n1} is set by the mean dark matter density. This acts as a lower bound because the origin of merger events such as GW190814 is most likely in dark matter dense environments (galactic halos) that imply a higher number density of primordial black holes (e.g., the mean dark matter density of a dark matter halo is $\sim 200$ times the mean density of dark matter we assume in \Eq{eq:n1}.

\begin{figure*}[ht]
\centering
\includegraphics[width=2.0\columnwidth]{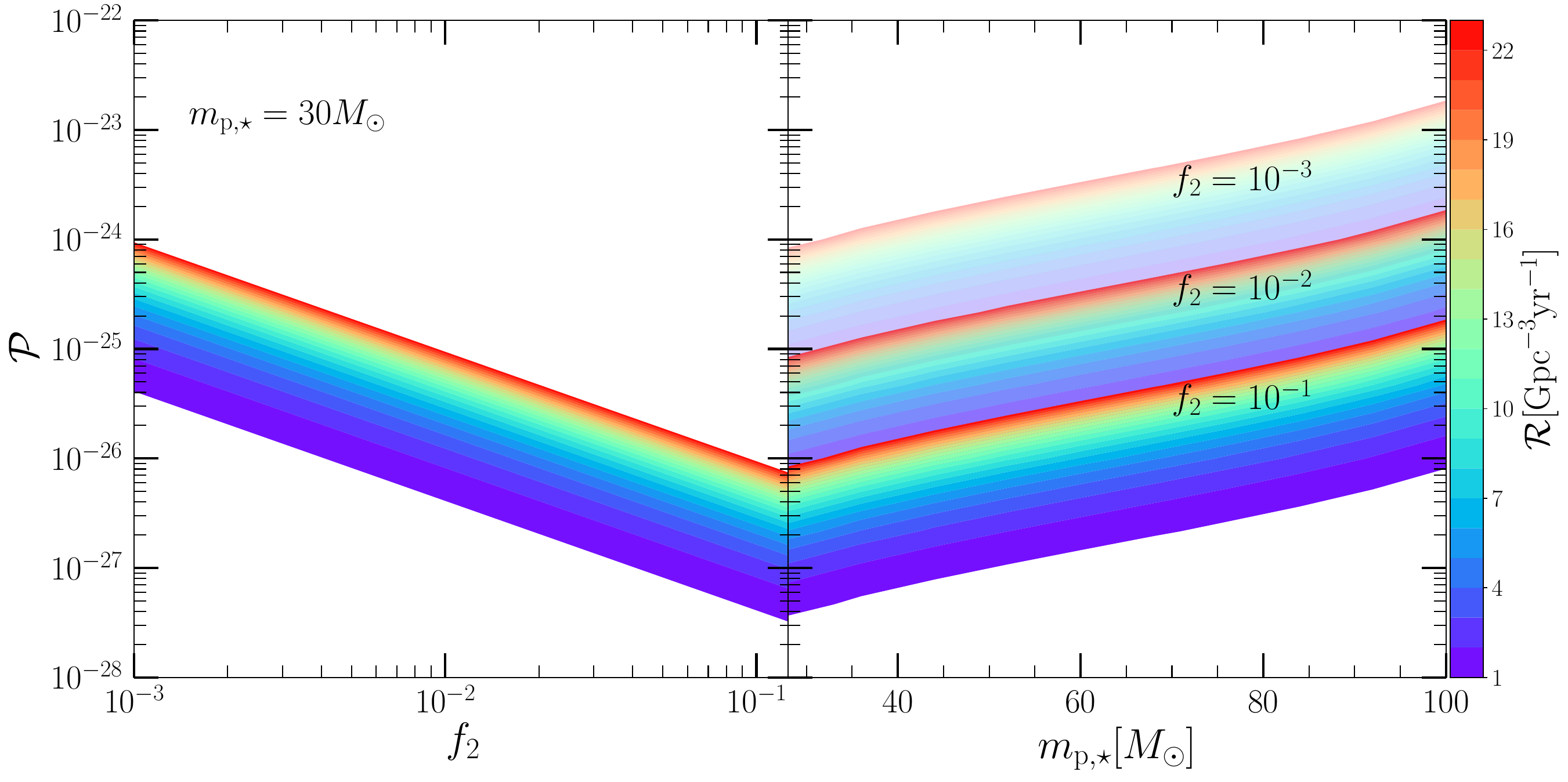} 
\caption{Event rate $\mathcal{R}$ of $2.6\Msun$ primordial black holes with $23\Msun$ stellar black hole in the $\mathcal{P} - f_2$ (left) and $\mathcal{P} - M_{p}^{\mathrm{min}} $ (right)  parameter space. The color coding corresponds to the derived rate from GW190814.}
\label{fig:fig2}
\end{figure*}

We can obtain the number density of $23 \Msun$ stellar origin black holes in the following way. Assume that the progenitors of such black holes are massive stars whose formation rate is given by the cosmic star formation rate, $\psi(z)$, obtained from observations of star forming galaxies out to high redshift \cite{Mandau2014}, 
\begin{equation} 
\psi(z) = 0.015 \frac{(1+z)^{2.7}}{1+[(1+z)/2.9]^{5.6}} \rm{M}_\odot \rm{yr}^{-1}\rm{Mpc}^{-3}.
\label{eq:SFR}
\end{equation}
The mass converted to stars of mass $m$ at redshift $z$ is distributed according to a Salpeter-like \cite{Salpeter1955} mass function rate density, 
\begin{equation} 
\xi ' (m,z) =  \frac{dN}{ dV \,dt \, d \ln m } \sim m^{-1.35}
\label{eq:xi}
\end{equation} 
We can normalize the  mass function at each redshift by requiring that the integral of the rate density of formation over all stellar masses is given by the star formation rate, i.e, $\psi(z)  = \int \xi ' (m,z) \,dm$, where the limits of integration are from $0.08 \Msun$ to $120 \Msun$. We assume conservatively that the mass function is independent of redshift (metallicity can change the slope of the Salpeter mass function) -- a shallower power law mass function on small scales (e.g., Kroupa \cite{2001MNRAS.322..231K}) increases the abundance of high mass stars if normalized the same way, and thus make the presented arguments even more stringent.
With this formalism, the number density rate of stellar progenitors with mass greater than $m_{{\rm p},\star}$ (and up to $m_{\mathrm{p,max}} = 120 \Msun$) is then
\begin{equation} 
\dot{n}_\star(z) = \int_{m_{{\rm p},\star}}^{m_{\mathrm{p,max}}} \frac{\xi ' (m,z)}{m} \, dm.
\label{eq:star}
\end{equation} 

We interpret $\dot{n}_\star$ as the rate of formation of stellar progenitors whose stellar remnant is a $23 \Msun$ black hole.This is a maximal assumption as every star whose mass is greater than  $23 \Msun$ will produce a $23 \Msun$ black hole. Under this assumption, the number density rate $\dot{n}_2$ of black holes of mass $23 \Msun$ available to merge with a primordial black hole is $\dot{n}_1 = \dot{n}_\star$. Figure~\ref{fig:fig2} shows the redshift dependence of the formation rate of $23\Msun$ black holes from heavier stellar progenitors. The shape of this function is set by the star formation rate, eq.~\ref{eq:SFR} (peaking at $z \approx 2$), while the amplitude of the function is set by the initial mass function of stellar masses, eq.~\ref{eq:xi}. Finally an important caveat is that only stars with metallicities less than $0.1 Z_\odot$ would be able to produce a black hole remnant of the required mass which subsequently reduces $\dot{n}_1$ \cite{2020ApJS..247...48K}.

\section{Results}
The LIGO observation of  GW190814 provides an estimate of the merger rate of $23\Msun$ black holes with $2.6\Msun$ compact objects. This observed rate is ${\cal R}_{\mathrm{obs}} = [1-23] \rm{Gpc}^{-3}\rm{yr}^{-1}$. In order to assess the probability that the $2.6\Msun$ object is a primordial black hole, we set the merger rate of \Eq{eq:merger_rate} equal to the observed rate, i.e., ${\cal R} = {\cal R}_{\mathrm{obs}}$, and then find the values of ${\cal P}$ and minimum progenitor mass $m_{{\rm p},\star}$ that can satisfy the equality. 

Figure~\ref{fig:fig2} shows the results of this calculation. We find that if all stars with masses greater than $30 \Msun$ produce $23 \Msun$ black holes then the probability of LIGO GW190814 being due to the merger of a $23 \Msun$ black hole with a $2.6 \Msun$ primordial black hole is between 
$ 10^{-27} < {\cal P} < 10^{-23}$
 for $10^{-3}\leq f_2\leq 10^{-1}$. The largest the $f_2$ the smaller  ${\cal P}$ is to maintain the same rate, while for larger $m_{\rm{p},\star}$ it needs to increase to accommodate the smaller number of stellar black holes available.

To interpret this result we need to characterize the physical meaning behind $\mathcal{P}$. The terms in \Eq{eq:merger_rate} (aside from $\mathcal{P}$) give all pairs of primordial and stellar black holes per volume per time. Therefore  $\mathcal{P}$ acts as a filter of how many of those black holes are in binaries, and of those how many would have merged per redshift interval. We can parametrize $\mathcal{P}$ as 
\begin{equation} 
\mathcal{P} = \left( \frac{N_\mathrm{binary}}{N_\mathrm{total}} \right) \left( \frac{t_u - \tau}{t_u} \right) 
\label{eq:P_parm}
\end{equation}
where $N_\mathrm{binary}/N_\mathrm{total}$ is the fraction of objects in binary systems, $t_u$ is the age of the universe, $\tau$ is the mean duration between the formation of the stellar black hole, the time to form the binary system and the time it takes for it to merge. In other words, the second term in eq.~\ref{eq:P_parm} quantifies how many of the binaries would have merged in the lifetime of the universe. A value of $\tau = 0$ means that merging is instantaneous after formation and all the binary systems would have merged. On the other hand a value of $\tau = t_u$ means that no binary system would have enough time to merge by today.

Let's consider the limiting case for which the timescale of the formation and merging of the binary is much less than the age of the universe and thus $\tau \sim 0$. 
The derived value of ${\cal P}$ in this case is just the fraction of primordial black holes in binaries needed to explain the observed rate. Here, the smallness of this number (${\cal P} \approx [10^{-27} - 10^{-23}]$) is extremely important -- it implies (unrealistically) that less than one primordial -- stellar mass black hole merger is needed in order to satisfy the requirements in the observed cosmological volume of LIGO, $V_{\mathrm{LIGO}}$. 

The observation of one event  requires {\it at least one} of all the primordial black holes in $V_{\mathrm{LIGO}}$ to have formed a binary. Therefore, $\cal P$ is limited by a minimum value  $\mathcal{P}_{\mathrm{min}} = 1/(n_2 V_{\mathrm{LIGO}})$, and as a consequence, \Eq{eq:merger_rate} limits the rate to a minimum value  $\mathcal{R}_{\mathrm{min}}$. This lower bound on the rate depends only on the formation rate of the stellar remnant partner and cannot be modified by adjusting parameters for the primordial black hole population. In other words, the minimum rate $\mathcal{R}_{\mathrm{min}}$ is set {\it only by the rate at which $23 \Msun$ black holes become available}.  

However there is an important caveat in this case. For example if $m_{\rm{p},\star} = 30 M_\odot$, $\mathcal{R}_{\mathrm{min}}$ is of order $10^5 \rm{Gpc}^{-3}\rm{yr}^{-1}$ (see \Fig{fig:fig2}) 
which would immediately exclude the possibility of the $2.6 \Msun$ being a primordial black hole since $\mathcal{R}_{\mathrm{min}}$ is 4-5 orders of magnitude larger than the observed rate. In order to get around this obstacle, these 5 orders of magnitude must be attributed to the probability of forming such a system realistically and not instantaneously  as previously assumed.  

In this particular example we can relax the assumption that   $\tau \sim 0$ by setting $(t_u - \tau)/t_u \sim 10^{-5}$, and  derive bound limits for the mean duration between forming a binary system and when the system merges, as $t_u \geqslant \tau \geqslant \tau_{\rm{min}}$, where $\tau_{\rm{min}} =  (1 - 10^{-5}) t_u$.  This lower limit, $\tau_{\rm{min}}$, is set by the value $m_{\rm{p},\star}$ because of the dependance of $\mathcal{R}_{\mathrm{min}}$ on the stellar black holes formation rate $\dot{n}_1$. The larger   $m_{\rm{p},\star}$ is, the fewer stars and thus stellar black holes are created, reducing $\mathcal{R}_{\mathrm{min}}$  and allowing for smaller values of $\tau$ while maintaining the observed rate. 

Note that $N_{\mathrm{binary}}/N_{\mathrm{total}}$ doesn't have to be necessarily at its minimum value; there may be more than one such binary system in  $V_{\mathrm{LIGO}}$. As $N_{\mathrm{binary}}/N_{\mathrm{total}}$ approaches its maximum value of $N_1/N_2$, the requirement to maintain the observed rate  has to be accounted for by reducing the temporal factor in $\mathcal{P}$. This argument further limits $\tau_{\rm{min}}$ to values even closer to the age of the universe. One way to relax that constraint would be by lowering the value of $N_1$ using the fact that only stars with metallicities less than $0.1 Z_\odot$ would be able to produce a black hole remnant of the required mass \cite{2020ApJS..247...48K}. However, given that $n_2$ is extremely large (the MW contains of order $10^{9}$ primordial black holes if we assume $f=0.01$) a reduction in  $n_1$ by even few orders of magnitude will have negligible effect on the results.

\section{Conclusions}

We investigated the plausibility of a primordial black hole origin of the secondary object in GW190814. We found that even if primordial black holes account for at most one percent of the dark matter in the universe, the abundance of  primordial black holes leads to an observed rate that highly exceeds the observed rate of such LIGO events. In other words, the large number of primordial black holes imply that {\it as long as stellar progenitors produce a $23\Msun$ black hole, it is guaranteed that at least one merger event will take place within a hubble time}. 

More specifically, we showed that if at least one merger event takes place between a primordial black hole and a stellar origin black hole within the LIGO volume implies that the time it takes for the formation of the stellar mass black hole, the capture to a binary and the subsequent inspiral and merger with a primordial black hole must be very close to the age of the universe, $\tau \ge \tau_{\rm{min}} =  (1-10^{-5}) t_u$. This is a hard bound, as any smaller value of $\tau_{\rm{min}}$ would give rise to a higher merger rate than what has been observed with LIGO. In other words, if such merger events can occur on faster timescales the merger rate will be higher than observed. However the observation of GW190814 suggests  that such a merger did take place on a timescale $t(z=0.053) \approx 12.97 {\rm Gyr} < \tau_{\rm{min}}$, contradicting our findings. 

Therefore to summarize, the large abundance of  ${\cal{O}}(20) M_\odot$ stellar origin black holes inferred from the LIGO merger events of such black holes, together with the observed redshift of GW190814 suggest that a primordial black hole origin of the secondary component of GW190814 is rather unlikely.

\section{Acknowledgments}
We thank the referee for numerous suggestions that improved the presentation of the manuscript. We benefitted from useful conversations with Manuel Buen-Abad, Jatan Buch, JiJi Fan, Leah Jenks, John Leung and Michael Toomey. We gratefully acknowledge the support of Brown University. 

\bibliography{manuscript}

\end{document}